\documentclass{moriond}

\def\be{\begin{equation}}
\def\ee{\end{equation}}
\def\bea{\begin{eqnarray}}
\def\eea{\end{eqnarray}}

\newcommand{\Photo}{\includegraphics[height=35mm]{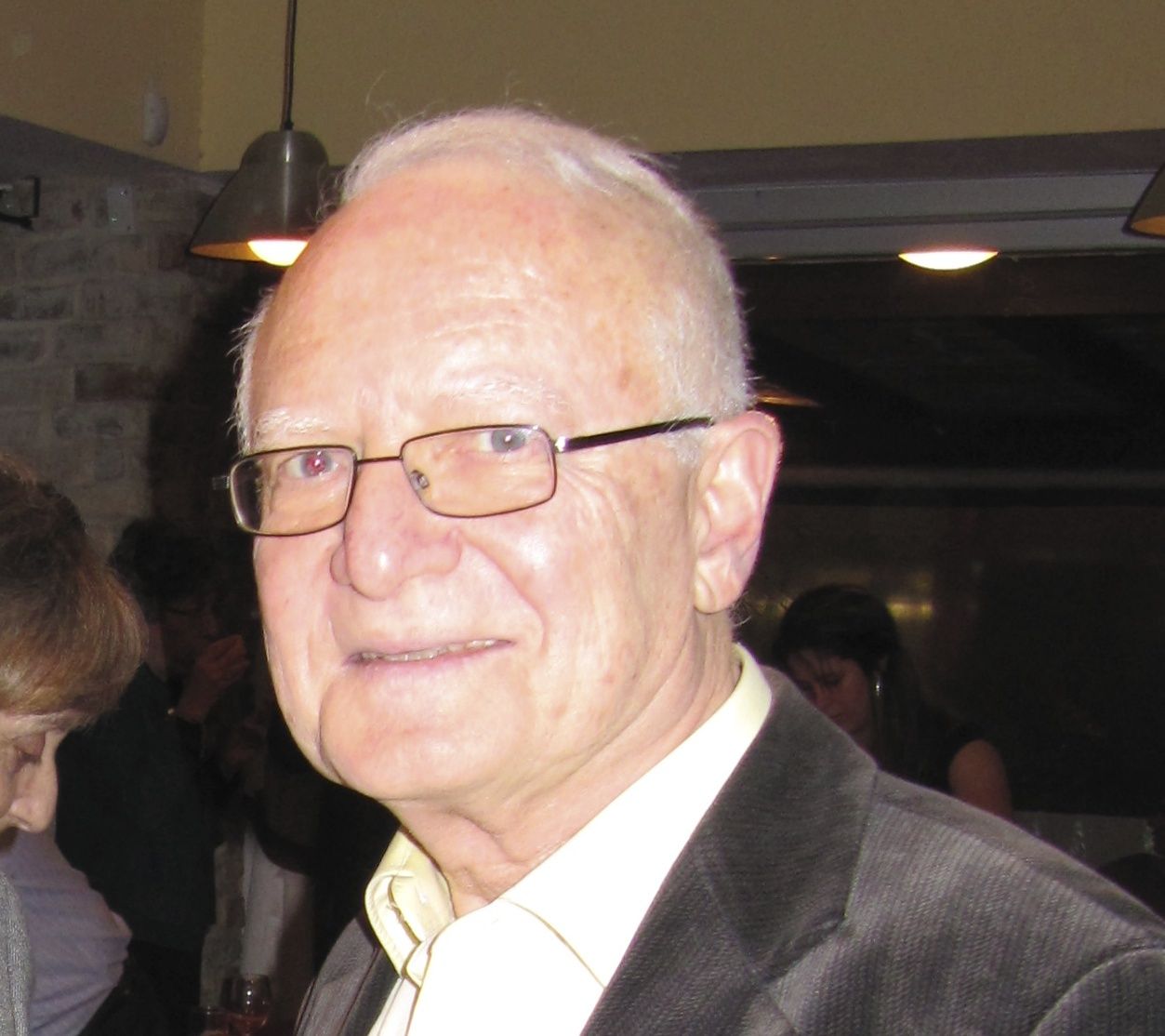}}
\usepackage{lineno}

\begin{document}
\vspace*{4cm}
\title{RECENT HERA RESULTS ON PROTON STRUCTURE}

\author{ AHARON LEVY }

\address{Tel Aviv University, Tel Aviv, Israel\\ On behalf of the H1 and ZEUS collaborations}

\maketitle\abstracts{
The latest results of the two HERA collaborations, H1 and ZEUS, are presented. They include the most recent  measurements of the longitudinal structure funcion $F_L$ from both collaborations. Also presented are high $Q^2$ measurements from the ZEUS collaboration in the high Bjorken $x$ region up to values of $x\cong 1$. 
}

\section{Introduction}

HERA was a high-energy electron\footnote{Here and in the following the term electron denotes generically both the electron and the positron.}-proton collider, at a centre-of-mass (cms) energy of 320 GeV. It started operating in 1992 and was closed in 2007.  Due to the accessible high values of virtuality, $Q^2$,  of the exchanged boson (see Fig. 1), reaching values up to about 40 000 GeV$^2$, it could 'look' into the proton with a resolution $\lambda$ of about 10$^{-3}$ fm.

\begin{figure}[h!]
\begin{minipage}{0.25\linewidth}
\centerline{\includegraphics[width=0.7\linewidth]{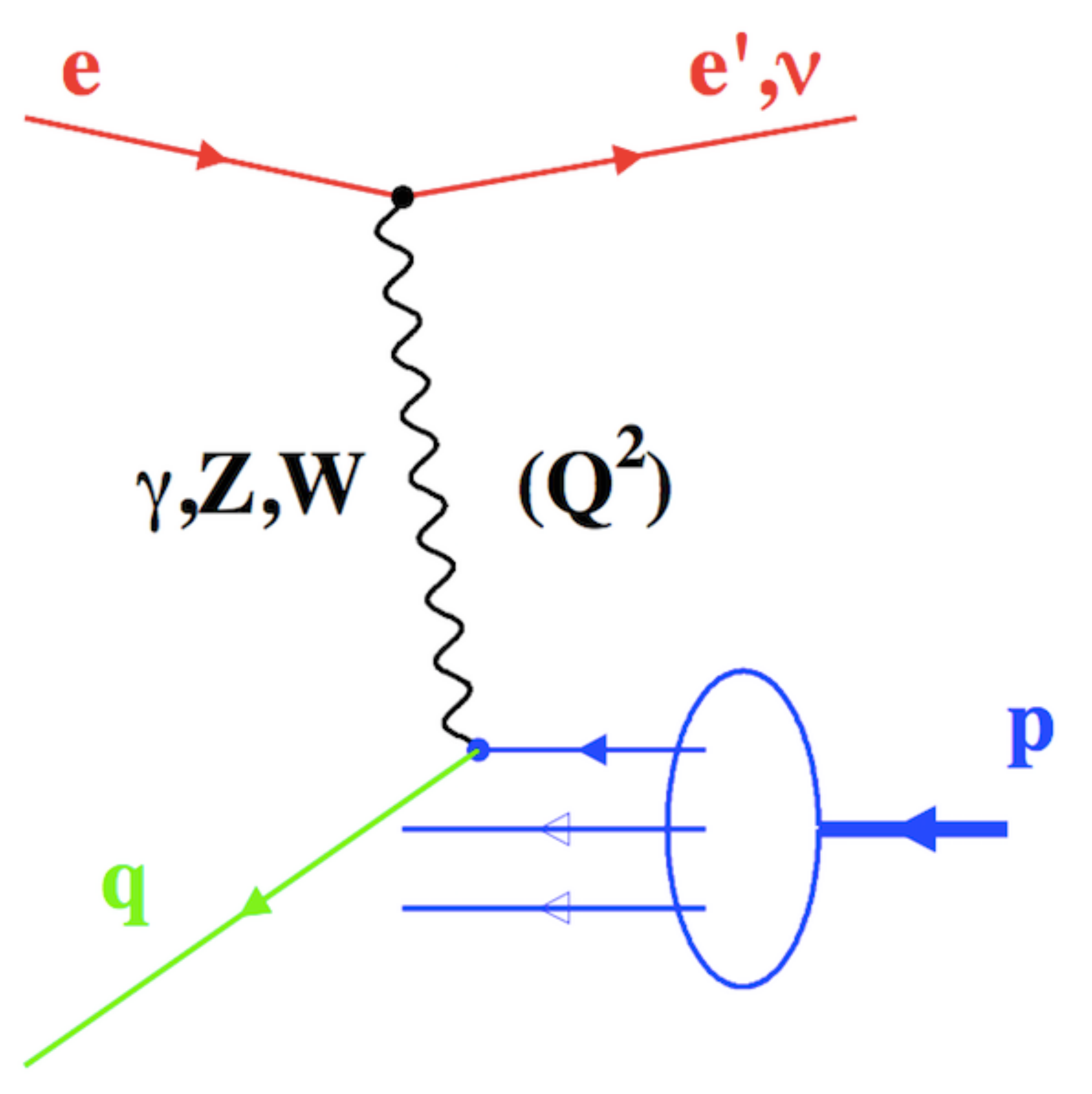}}
\vspace{-0.5cm}
\caption{Diagram describing $ep$ collisions.}
\end{minipage}
\hfill
\begin{minipage}{0.7\linewidth}
At HERA many experiments were performed by changing the virtuality of the exchanged photon from almost-real photons ($Q^2 \sim$ 0), the photoproduction region, through the start of the deep inelastic scattering (DIS) region, $Q^2 \sim$ 4 GeV$^2$ ($\lambda$=0.1 fm), to the very high-$Q^2$ region, $Q^2 \sim$ 40 000 GeV$^2$ ($\lambda=10^{-3}$ fm), where electroweak physics could be studied.
\end{minipage}
\end{figure}

In this talk, two of the most recent results concerning the proton structure will be presented. The first is a measurement~\cite{h1fl,zeusfl}, by both collaborations, of the longitudinal structure funcion, $F_L$. The second, carried out by the ZEUS collaboration~\cite{zeushighx}, is the high $Q^2$ measurements in the high Bjorken $x$ region up to values of $x\cong 1$.

\section{Measuring the longitudinal structure function $F_L$}

The $F_L$ structure function was measured at HERA only during the last months of its running in 2007. Up to that time, measurements of the $F_2$ structure function were limited~\cite{rmp} to low $y$, where $y$ is the fraction of the lepton energy transferred to the proton in its rest frame. The coefficient in front of the $F_L$ term is $y^2$ and thus its contribution to the cross section, compared to that of the $F_2$ structure function, is very small for low-$y$ values. 

\begin{figure}[h!]
\begin{minipage}{0.38\linewidth}
\includegraphics[width=0.95\linewidth]{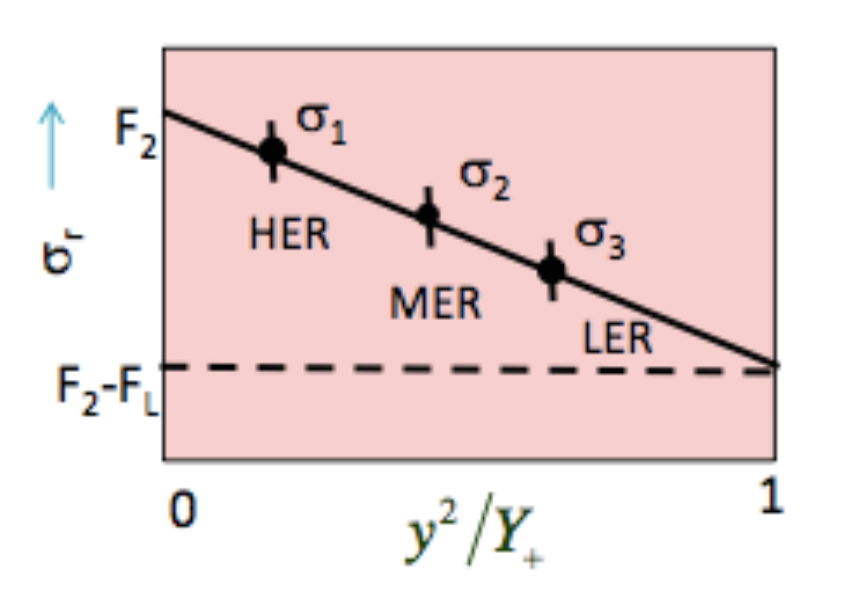}
\vspace{-0.5cm}
\caption{A sketch of the linear dependence of $\sigma_r$ on $y^2/Y_+$. The intercept is $F_2$ and the slope gives $F_L$.}
\label{fig:Rosen}
\end{minipage}
\hfill
\begin{minipage}{0.6\linewidth}
The reduced cross section, $\sigma_r$, can be expressed by two terms in the region where the $Z$ exchange can be neglected, meaning $Q^2$ values far below the square of the $Z$ mass,
\begin{equation}
\sigma_r = F_2(x,Q^2) - (y^2/Y_+)F_L(x,Q^2),
\end{equation}
where  $Y_+ = 1 + (1 - y)^2$. Measuring $\sigma_r$ at different $y$ but at the same $x,Q^2$ values gives a linear dependence of $\sigma_r$ on $y^2/Y_+$ and therefore allows a simultaneous determination of the two structure functions $F_2$ and $F_L$. This is shown in Fig.~\ref{fig:Rosen}. Since $y = Q^2 / (x s)$, where $s$ is the cms squared of the $ep$ system, the way to vary $y$ is to vary $s$. This has been done by changing the proton-beam energy to 460 and 575 GeV.
\end{minipage}
\end{figure}

The determination of $F_L$ needs the measurement of high-$y$ events. The variable $y$ is a function of the scattered electron kinematics, 
\begin{equation}
y = 1 - \frac{E^\prime}{E_e (1 - \cos\theta)},
\end{equation}
where $E_e$ is the electron-beam energy, $E^\prime$ and $\theta$ are the energy and angle of the scattered electron, respectively. Thus high values of $y$ means low $E^\prime$ of the scattered electron. Electron finders of both collaborations, prior to this measurement, were very well trained to identify scattered electrons with energies $E^\prime >$ 10 GeV. For lower energies, the efficiencies and purities of the finders deteriorate because of the photoproduction background. The ZEUS collaboration succeeded to improve their finder to allow  to include in the $F_L$ measurements events with $E^\prime >$ 6 GeV. The H1 collaboration, whose detector is better suited for this measurement could go down to $E^\prime >$ 3 GeV. 
\begin{figure}[h!]
\begin{minipage}{0.45\linewidth}
\includegraphics[width=0.55\linewidth]{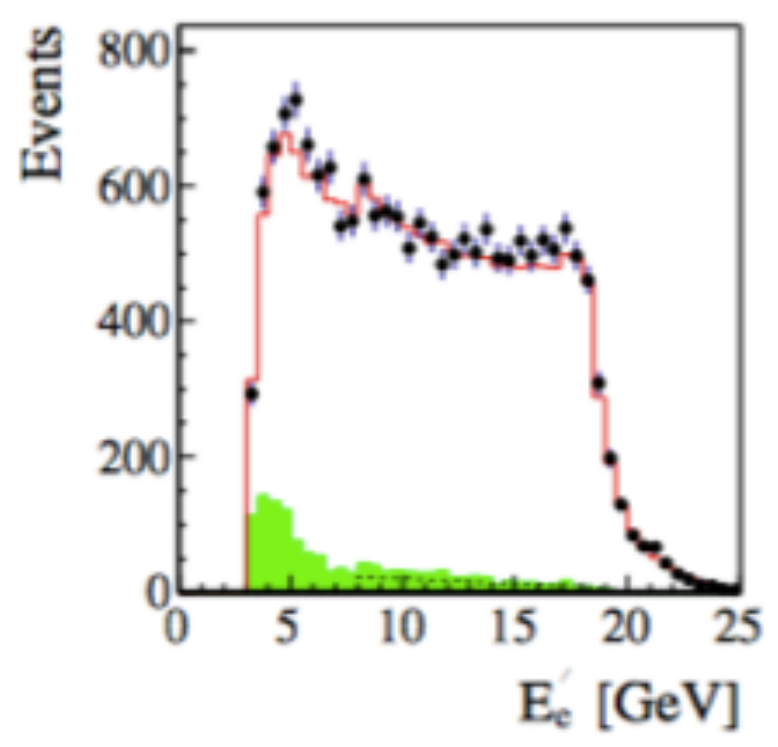}
\caption{Comparison of data and Monte Carlo for the scattered electron energy distribution at proton-beam energy of 460GeV for the H1 collaboration. The shaded region is the photoproduction background.}
\label{fig:ee-H1}
\end{minipage}
\hfill
\begin{minipage}{0.45\linewidth}
\includegraphics[width=0.55\linewidth]{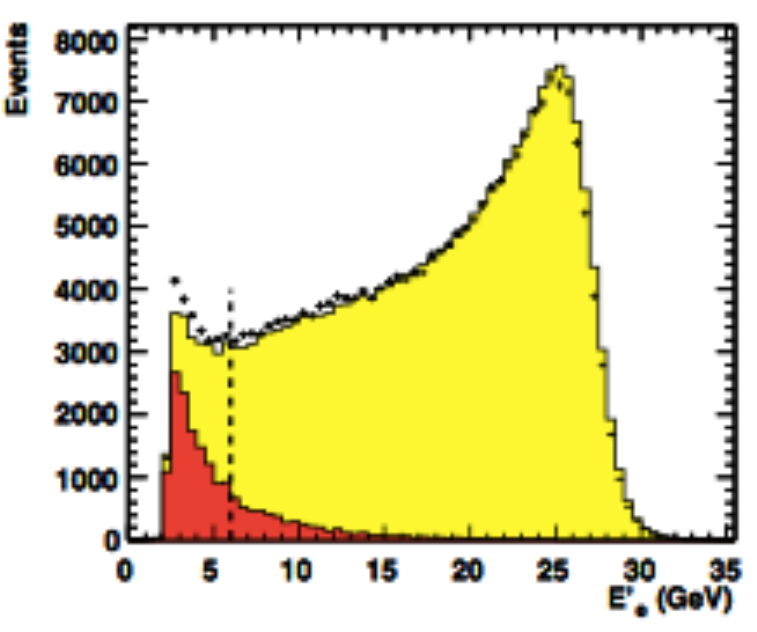}
\caption{Comparison of data and Monte Carlo for the scattered electron energy distribution at proton-beam energy of 460GeV for the ZEUS collaboration. The dark-shaded region is the photoproduction background.}
\label{fig:ee-ZEUS}
\end{minipage}
\end{figure}
Control plots showing a comparison between data and Monte Carlo for the $E^\prime$ variable for the low-energy run (proton beam of 460 GeV) are shown in Figs.~\ref{fig:ee-H1} and~\ref{fig:ee-ZEUS}. The photoproduction background is shown in the dark-shaded region and is seen to increase sharply for low $E^\prime$ values.

\begin{figure}[h!]
\begin{minipage}{0.5\linewidth}
Following the limitations on the energy of the scattered electron, the ZEUS collaboration measured $F_L$ in the kinematic range $9 < Q^2 < 110$ GeV$^2$ while the H1 collaboration covered the region $1.5 < Q^2 < 800$ GeV$^2$. The results are shown in Fig.~\ref{fig:fl-both}. The uncertainties of the ZEUS results are larger than those of H1. The ZEUS results, though consistently lower than those of H1, are consistent with them because of the correlated uncertainties. Taking into account the correlations between the ZEUS data points and neglecting the correlations between the H1 data points a $\chi^2$ of 12.2 is obtained for 8 degrees of freedom. The predictions shown by the shaded area are in reasonable agreement with both data sets.
\end{minipage}
\hfill
\begin{minipage}{0.48\linewidth}
\includegraphics[width=0.98\linewidth]{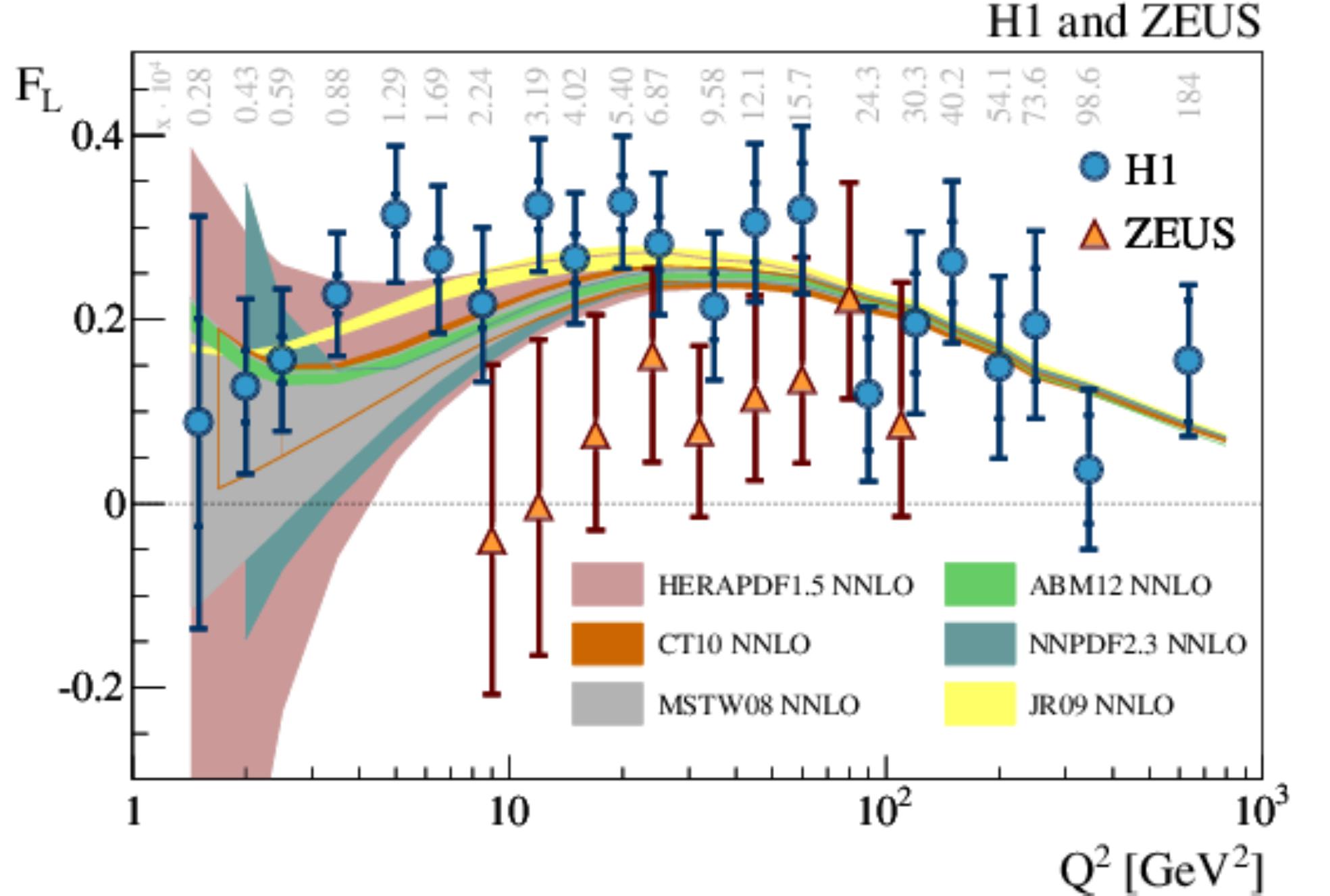} 
\caption{$F_L$ as a function of $Q^2$ as measured by the H1 and ZEUS collaborations. The shaded area are predictions based on different parameterisation, as indicated in the figure.} 
\label{fig:fl-both} 
\end{minipage}
\end{figure}

\section{High x, extending to $x\cong 1$}

\begin{figure}[h!]
\begin{minipage}{0.38\linewidth}
\includegraphics[width=0.95\linewidth]{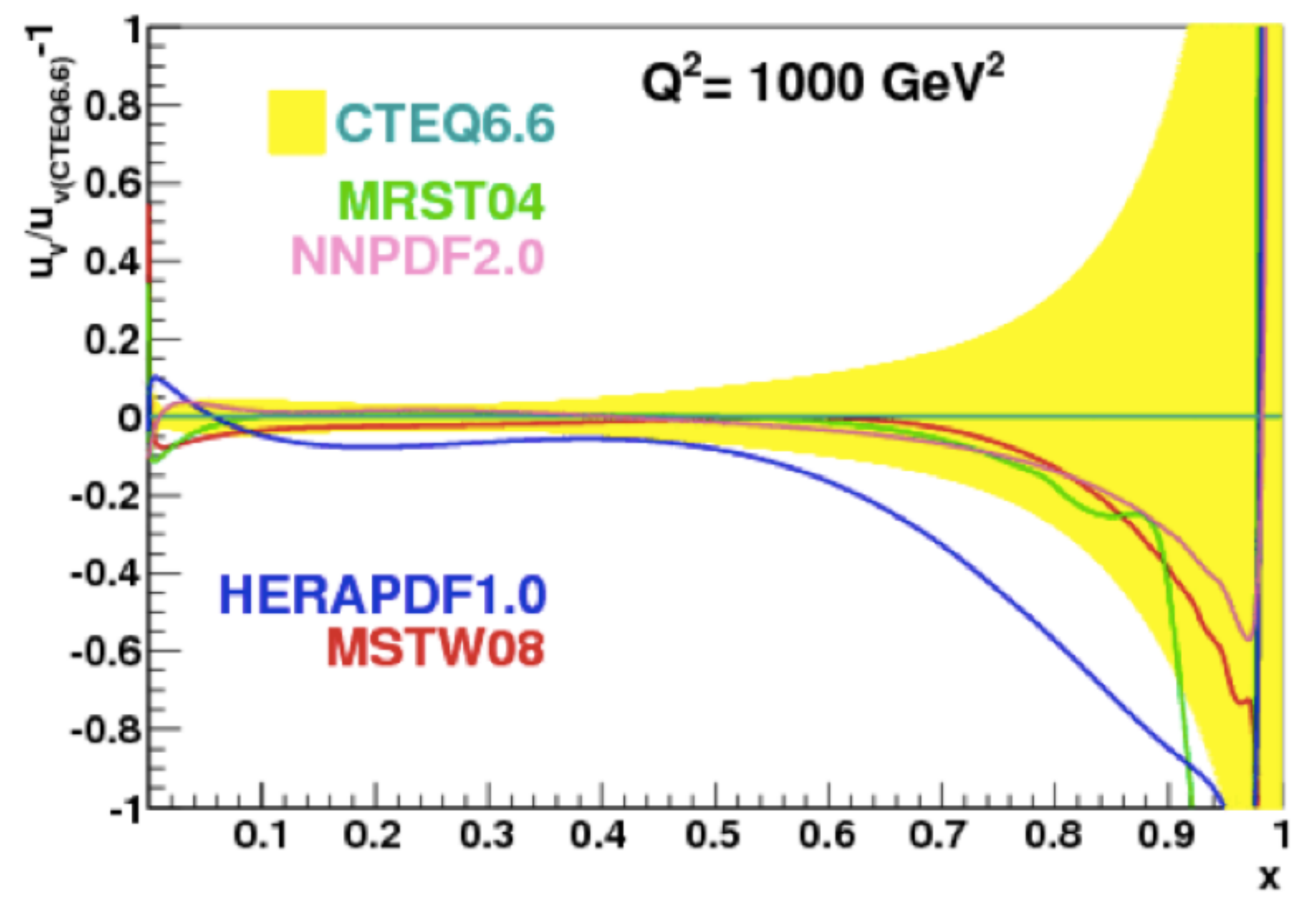}
\vspace{-0.5cm}
\caption{Example of the sizable differences between some parameterisation description of the $u$ valence quark, $u_V$.}
\label{fig:motivation}
\end{minipage}
\hfill
\begin{minipage}{0.6\linewidth}
The DIS cross sections have been measured by both collaborations with very high precision. These measurements were combined and produced text-book results with even higher precision~\cite{combined}. Nevertheless, the highest $x$ value for which measurements were done was 0.65. There are fixed-target experiments~\cite{pl:b223:485,pl:b282:475,jferson} which measure higher values of $x$ but in a low $Q^2$ region. In global perturbative quantum chromodynamic fits of parton distribution functions (PDFs), a parameterisation of the form $(1 - x)^\beta$ is assumed in order to extend PDFs ro $x$ = 1. Although all fitters use the same parameterisation, sizeable differences are obtained in the high-$x$ region~\cite{allen-eps}, as shown in Fig.~\ref{fig:motivation}.
\end{minipage}
\end{figure}

\begin{figure}[h!]
\begin{center}
\includegraphics[width=0.75\linewidth]{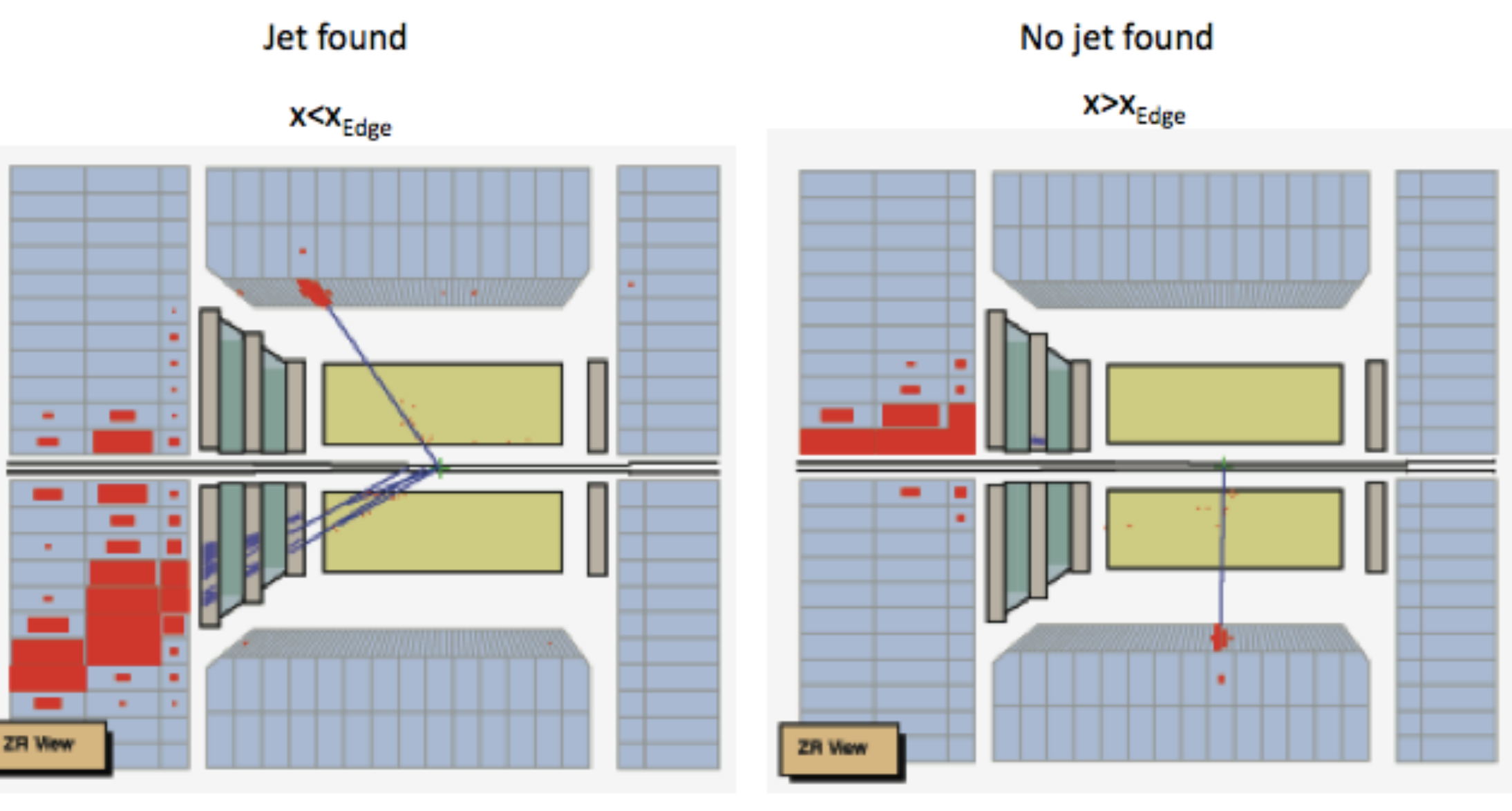}
\caption{Left-hand side: a one-jet event with a scattered electron in the BCAL and the jet fully contained in the FCAL.Also seen in FCAL are the proton remnent. Right-hand side: A zero-jet event where the scattereed electron is in BCAL and the jet remains inside the beam pipe. The proton remnant and possibly some energy emerging from the jet in the beampipe are seen in FCAL.}
\label{fig:xedge}
\end{center}
\end{figure}

The ZEUS collaboration showed in an earlier publication~\cite{epj:c49:523-544} that the kinematics of HERA and the design of the detectors allow extension of the measurements of the neutral current (NC) cross sections up to $x$ = 1. The results presented here are based on a much larger data sample and an improved analysis procedure.

A typical NC high-$Q^{2}$ and high-$x$ event consists of the scattered electron and a high-energy collimated jet of particles in the direction of the  struck quark. The electron and the jet are balanced in transverse momentum. The proton remnant mostly disappears down the beam pipe. The $x$ and $Q^2$ of events, in which the jet is well contained in the detector, may be determined by various techniques. However,  the maximum $x$ value that can be reached is limited by the fact that at the low values of $y$ typical of these events, the uncertainty on $x=Q^2/ys$  increases as $\Delta x\sim \Delta y/y^2$. An improved $x$ reconstruction is achieved by observing that, in the limit of $x\rightarrow 1$, the energy of the struck quark represented by a collimated jet is $E_\mathrm{jet} \cong xE_p$. The  expression for $x$ is 
\begin{equation}
x = \frac {E_\mathrm{jet}(1+\cos \theta_\mathrm{jet})}{2 E_p \left( 1- \frac {E_\mathrm{jet}(1-\cos\theta_\mathrm{jet})}{2E_{e}} \right) } \, ,
\label{eq-xpt}
\end{equation}
where $\theta_\mathrm{jet}$ is the scattering angle of the jet in the detector. 

As $x$ increases and the jet associated with the struck quark disappears down the beam-pipe (see Fig.~\ref{fig:xedge}), the ability to reconstruct $x$ is limited by the energy loss. However, in these events, the cross section integrated  from a certain limit in $x$, $x_\mathrm{edge}$, up to $x=1$ is extracted. The value of $x_\mathrm{edge}$  below which the jet is fully contained in the detector depends on $Q^2$ and the higher the $Q^2$, the higher the value of $x_\mathrm{edge}$.

\begin{figure}[h!]
\begin{minipage}{0.48\linewidth}
\includegraphics[width=0.95\linewidth]{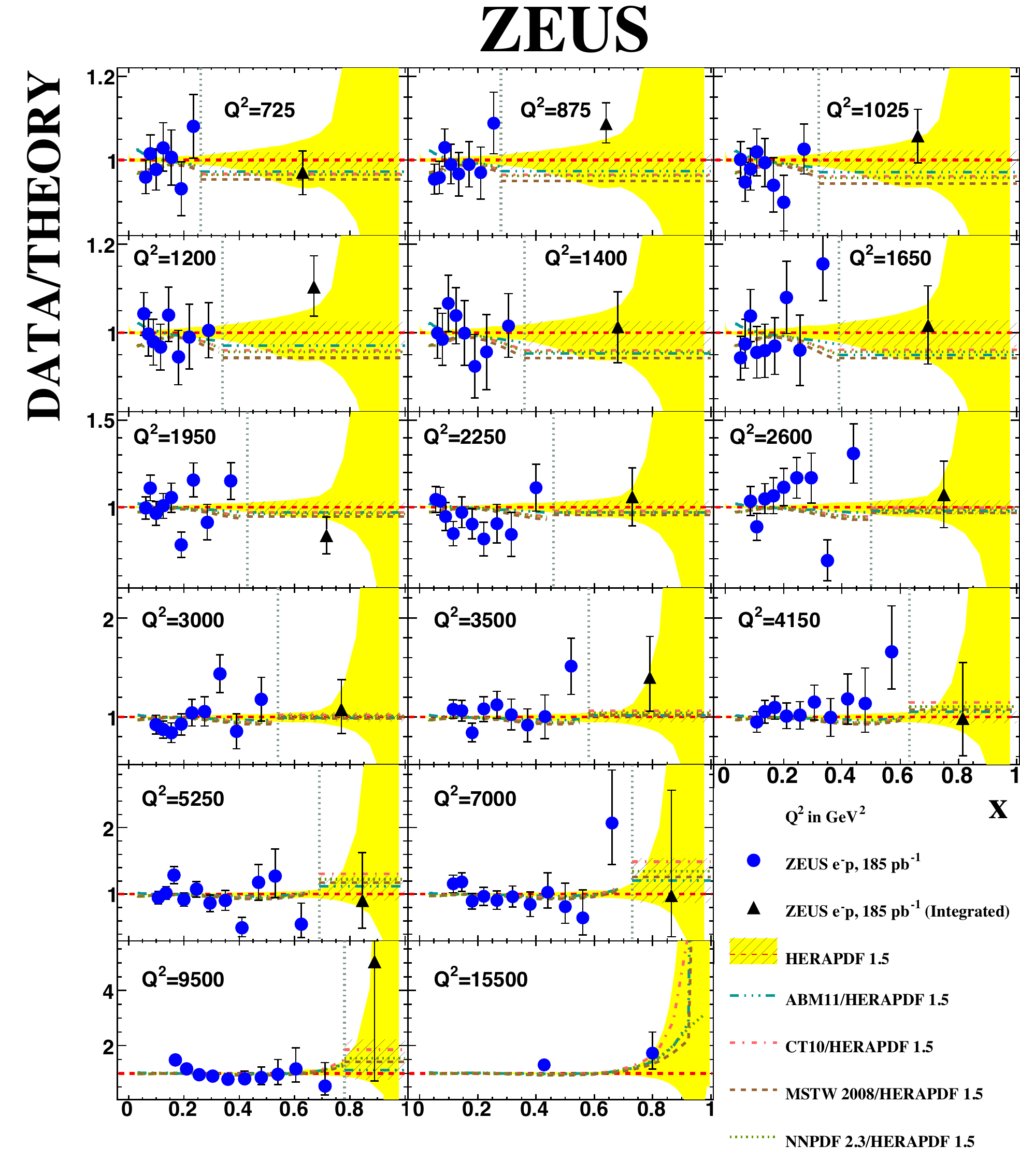}
\caption{
Ratio of the double-differential cross section for NC $e^-p$ 
scattering and  of the double-differential
cross section integrated over $x$ 
to the Standard Model expectation evaluated using the HERAPDF1.5 PDFs as a function of $x$ at different $Q^2$ values as described in the legend.  For HERAPDF1.5, the uncertainty is given as a band. The expectation for the integrated bin is also shown as a hatched box. 
The error bars show the statistical and systematic uncertainties added in quadrature.
The expectations of other commonly used PDF sets normalised to HERAPDF1.5 PDFs are also shown, as listed in the legend. Note that the scale on the $y$ axis changes with $Q^2$. }
\label{fig:eMp}
\end{minipage}
\hfill
\begin{minipage}{0.48\linewidth}
\includegraphics[width=0.95\linewidth]{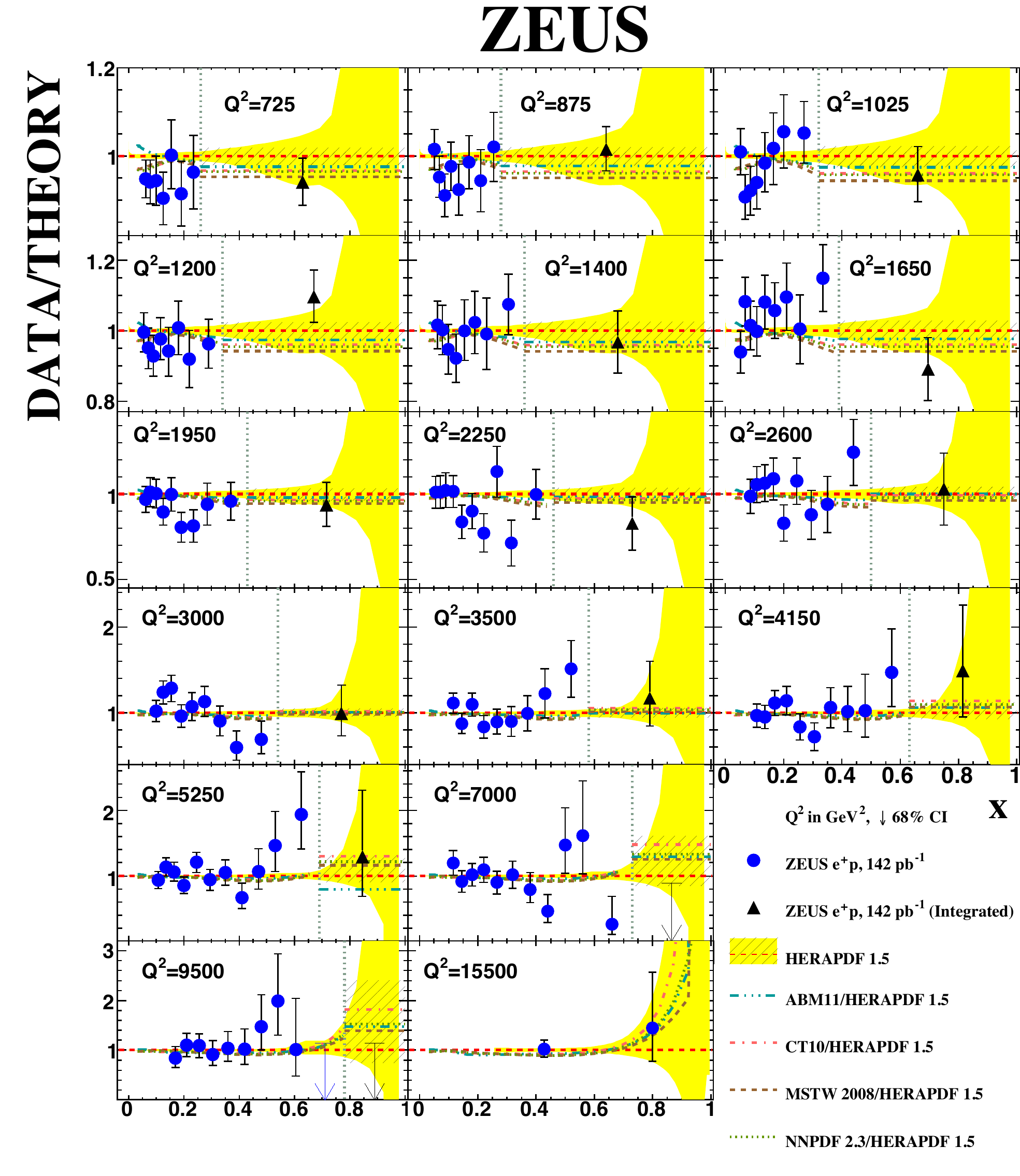}
\caption{
Ratio of the double-differential cross section for NC $e^+p$ 
scattering and  of the double-differential
cross section integrated over $x$ 
to the Standard Model expectation evaluated using the HERAPDF1.5 PDFs as a function of $x$ at different $Q^2$ values as described in the legend.  For HERAPDF1.5, the uncertainty is given as a band. The expectation for the integrated bin is also shown as a hatched box. 
The error bars show the statistical and systematic uncertainties added in quadrature.
The expectations of other commonly used PDF sets normalised to HERAPDF1.5 PDFs are also shown, as listed in the legend. Note that the scale on the $y$ axis changes with $Q^2$. }
\label{fig:ePp}
\end{minipage}
\end{figure}

The double-differential Born-level cross sections as a function of $Q^2$ and $x$ have been measured in finer binning in $x$ because of the large data samples in this analysis (53 099 for the $e^- p$ and 37 361 for the $e^+ p$ sample).  For the highest integrated $x$ bin, the respective average cross sections, defined as
\begin{equation}
I(x) = \frac{1}{1-x_{\rm {edge}}}\int_{x_{\rm {edge}}}^{1}\frac{d^2\sigma(x,Q^2)}{dxdQ^2}dx
\;\; ,
\label{eqn-I(x)}
\end{equation}
have been obtained and plotted at $x=(x_\mathrm{edge}+1)/2$. The ratio of the measured cross sections to those expected from HERAPDF1.5~\cite{herapdf1.5} are shown in Figs.~\ref{fig:eMp} and~\ref{fig:ePp}.  Note that for bins where no events are observed, the limit is quoted at $68$\% probability,  neglecting the systematic uncertainty. Also shown are the predictions from a number of other PDF sets (ABM11~\cite{abm11}, CT10~\cite{ct10}, MSTW2008~\cite{mstw2008}, NNPDF2.3~\cite{nnpdf2.3}), normalised to the predictions from HERAPDF1.5.  Within the quoted uncertainties, the agreement between measurements and expectations is good.

\section{Summary}

Final measurements of  the $F_L$ structure functions are being published by HERA. The H1 collaboration covers a large kinematic range in $Q^2$, $1.5 < Q^2 < 800$ GeV$^2$. This is made possible by measuring scattered electrons down to 3 GeV due to good tracking and electromagnetic calorimetry in the rear direction. The results of the ZEUS collaboration in the $Q^2$ region covered by their measurements, $9 < Q^2 < 110$ GeV$^2$, are in general lower that those of H1 but taking into account correlated uncertainties, are consistent with those of H1. Both results are consistent with expectations, though at low $Q^2$ there are large uncertainties in the theoretical predictions.

The ZEUS collaboration measured double-differential cross sections for $e^\pm p$ NC DIS events at $Q^2 >$ 725 GeV$^2$ up to $x\cong 1$. Fine binning in $x$ and extension of kinematic coverage up to $x\cong 1$ make the data important input to fits constraining the PDFs in the valence-quark domain.

\section*{Acknowledgments}

This activity was partially supported by the Israel Science Foundation.


\section*{References}

\begin{thebibliography}{99}
\bibitem{h1fl} H1 Collaboration, V. Andreev et al., Eur. Phys. J. {\bf C 74} (2014) 2814.
\bibitem{zeusfl}ZEUS Collaboration, H. Abramowicz et al., DESY-14-053.
\bibitem{zeushighx} ZEUS Collaboration, H. Abramowicz et al., Phs. Rev. {\bf D 89} (2014) 072007.
\bibitem{rmp} See e.g. H. Abramowicz and A. Caldwell, Rev. Mod. Phys. {\bf 71} (1999) 1275.
\bibitem{combined} H1 and ZEUS Collaborations, F.D. Aaron et al., JHEP {\bf 1001} (2010) 109.
\bibitem{pl:b223:485}
BCDMS Collaboration, A.C. Benvenuti et al., Phys. Lett. {\bf B 223} (1989) 485.
\bibitem{pl:b282:475}
L.W. Whitlow et al., Phys. Lett. {\bf B 282}  (1992) 475.
\bibitem{jferson} S.P. Malace et al., Phys. Rev. {\bf C 80}  (2009) 035207.
\bibitem{allen-eps}A. Caldwell for the ZEUS Collaboration, {\it Measurement of positron-proton neutral current cross sections at high Bjorken-x with the ZEUS detector at HERA,} presented at the EPS2013, Stokholm, July 2013.
\bibitem{epj:c49:523-544}
ZEUS Coll., S. Chekanov et al., Eur. Phys. J. {\bf C 49} (2007) 523.


\bibitem{herapdf1.5}
V. Radescu, {\it Combination of QCD Analysis of the HERA Inclusive Cross Sections}, arXiv: 1308.0374 [hep-ex] (2013).


\bibitem{abm11}
S. Alekhin, J. Bl\"umlein and S.O. Moch, PoS {\bf LL2012}  (2012) 016.

\bibitem{ct10}
M. Guzzi et al., {\it CT10 parton distributions and other developments in the global QCD analysis}, SMU-HEP-10-11 (2011), arXiv: 1101.0561 [hep-ph] (2011).

\bibitem{mstw2008}
A.D. Martin et al., Eur. Phys. J. {\bf C 63}  (2009) 189.

\bibitem{nnpdf2.3}
R. Ball et al., JHEP {\bf 1304}  (2013) 125.

\end{thebibliography}
\end{document}